\RequirePackage{fix-cm}
\documentclass{svjour3mod}                     
\smartqed  
\usepackage{graphicx}
\usepackage{amsfonts}
\usepackage{mathtools}
\newcommand{\ket}[1]{|{#1}\rangle}

\newcommand{\bra}[1]{\langle{#1}|}

\newcommand{\exval}[1]{\langle{#1}\rangle}

\newcommand{\refx}[1]{Eq.~(\ref{#1})}

\begin{document}

\title{Getting information via a quantum measurement: the role of decoherence
}

\titlerunning{Decoherence and Quantum Measurements}        

\author{ Pietro Liuzzo-Scorpo \and Alessandro Cuccoli \and  Paola Verrucchi 
}


\institute{Pietro Liuzzo-Scorpo \at 
Dipartimento di Fisica Universit\`a di Firenze, Via G. Sansone 1,
I-50019 Sesto Fiorentino (FI), Italy, and School of Mathematical 
Sciences, The University of Nottingham, University Park, Nottingham NG7 
2RD, UK \and 
Alessandro Cuccoli \at
Dipartimento di Fisica Universit\`a di Firenze, and INFN 
Sezione di Firenze, Via G. Sansone 1, I-50019 Sesto Fiorentino (FI), 
Italy \and
Paola Verrucchi \at Istituto dei Sistemi Complessi ISC-CNR, Dipartimento 
di Fisica Universit\`a di Firenze, and INFN Sezione di Firenze, Via G. 
Sansone 1, I-50019 Sesto Fiorentino (FI), Italy
}

\date{} 

\maketitle

\begin{abstract}
In this work we investigate the relation between quantum measurements 
and decoherence, in order to formally express the necessity of the 
latter for obtaining an informative output from the former.
To this aim, we analyse the dynamical behaviour of a particular model, 
which is often adopted in the literature for describing projector 
valued measures of discrete observables. The analysis is developed by 
a recently introduced method for studying open quantum systems, namely 
the parametric representation with environmental coherent states: this 
method allows us to determine a necessary
condition that the evolved quantum state of the apparatus must fulfil
in order to have the properties that a measurement scheme requests it to 
feature. We find that this condition strictly implies decoherence in the 
system object of the measurement, with respect to the eigenstates of the 
hermitian operator that represents the measured observable.
The relevance of dynamical entanglement generation is highlighted, and 
consequences of the possible macroscopic structure of the 
measurement apparatus are also commented upon.
\end{abstract}

\section{Introduction}
\label{s.Introduction}

The profound relation between the quantum measurement process and 
decoherence is nowadays recognised as a key feature of quantum 
mechanics, not only from a foundational viewpoint but also when 
designing theoretical models or experimental setups aimed at capturing 
genuinely quantum behaviours of physical systems
\cite{BuschLM96,NamikiPascazioNakazato1997,Zurek2003,Schlosshauer}. 
However, in the formal construction of such relations there are still 
unclear 
points, that 
enforce the introduction of otherwise unnecessary concepts or 
even of additional axioms. It is not too stretched to say that 
these unclear points tend to nest where the crossover towards 
a macroscopic measurement apparatus comes into play 
and the quantum-to-classical transition consequently bursts into the 
description\cite{BuschLM96,Schlosshauer,GhirardiRW85,GhirardiRW86,NakazatoP93pra,Mermin98}. What makes it particularly problematic the 
formal treatment of such transition,
in the specific case of the measurement process, is 
the fact that it must 
exclusively concern the apparatus without affecting the object 
of the measurement, hereafter dubbed {\it principal 
system}, whose quantum character is not at issue. 
The aim of this work is that of giving a formal content to the role 
played by decoherence in the way we effectively probe the quantum world.

In the approach to which we will essentially refer, the measurement process is 
represented as an 
inherently dynamical one, entailing the definition of what is\par
\noindent
-{\it before} (the principal system in the state
about which we want to acquire information, and the
apparatus initialised in some dumb configuration),\par
\noindent
-{\it during} (the evolution ruled by the measurement coupling that 
generates entanglement between principal system and apparatus),\par
\noindent
- and {\it after} (the principal system in some final state and the 
apparatus in an informative and readable output configuration).
\par\noindent
Notice that the above splitting implies the possibility of switching
on/off the measurement coupling, a task which is most often accomplished
by reducing/increasing, respectively, the distance between
systems, with the implicit assumption that only short range interactions 
are relevant. Consistently, in this work we will not consider long range 
interactions, such as the Coulomb or gravitational ones.

Collecting clues from the above reflections we propose a 
description of a specific case of quantum 
measurement process in terms of the dynamical evolution of an Open 
Quantum System (OQS)\cite{Kraus83,Breuer2002,Holevo12,Rivas2012} whose 
environment is the measuring apparatus. We resort to a recently introduced 
method\cite{CalvaniEtal2013b} for 
studying OQS, namely the Parametric Representation with Environmental 
Coherent States (PRECS), which is specifically 
tailored to follow the environmental quantum-to-classical crossover.
Indeed one of the main feature of the PRECS is that of allowing an 
exact, and yet essentially asymmetric description of principal system 
and environment, with the former given in terms of parametrised pure 
states, and the latter strongly characterised by the use of generalised 
coherent states. 

The structure of the paper is as follows:
In Sec.\ref{s.Standard} we define the specific type of quantum 
measurement that will be considered, and introduce the model adopted for 
describing the process in the framework of the OQS
dynamics. The 
resulting evolution is studied in Sec.\ref{s.PRECS} by the PRECS, which 
is briefly reviewed and commented upon in this same section. The crucial 
point of how information about the principal system becomes available 
through the apparatus is finally tackled in Sec.\ref{s.ExtractInfo}, 
where decoherence appears as a necessary phenomenon in order for the 
measurement process to produce an informative output. Results are 
commented upon and conclusions drawn in Sec.\ref{s.Conclusions}.

\section{Standard model for unitary pre-measurements of discrete sharp 
observables}
\label{s.Standard}
Providing a formal description of the quantum measurement process is a 
challenge that, despite having been extensively taken on in the last 
century, cannot yet be considered definitively overcome. Different 
approaches have been proposed but no general consensus in favour of any 
of them has been 
reached (see Refs.\cite{BuschLM96} and \cite{Schlosshauer} 
for discussions and bibliographies on the subject). Aim of this 
Section is that of sketching the formalism we will 
refer to\cite{BuschLM96}, and define the specific case we will 
explicitly study.

Be $\Gamma$ the principal system, object of the measurement, and $\Xi$ 
its environment, acting as measuring apparatus: both systems are
described as quantum ones, with separable Hilbert spaces ${\cal 
H}_\Gamma$ and ${\cal 
H}_\Xi$, respectively. The composite system $\Psi=\Gamma+\Xi$, with 
separable Hilbert space ${\cal H}_\Psi={\cal H}_\Gamma\otimes{\cal 
H}_\Xi$, is assumed isolated: its state is therefore
pure at any time, $\rho_\Psi(t)=\ket{\Psi(t)}\bra{\Psi(t)}~\forall t$, 
and further presumed separable before the measurement starts
\begin{equation}
\ket{\Psi(t\le 0)}=\ket{\Gamma}\otimes\ket{\Xi}~;
\label{e.Psi0}
\end{equation}
notice that the validity of these assumptions should not be taken for 
granted, as extensively discussed, for instance, in 
Refs.\cite{BuschLM96,Schlosshauer}. 
The subsystems $\Gamma$ and $\Xi$ are certainly not 
isolated for $t>0$, and their respective state is $\rho_\Gamma(t)={\rm Tr}_\Xi
[\ket{\Psi(t)}\bra{\Psi(t)}]$, $\rho_\Xi(t)={\rm Tr}_\Gamma
[\ket{\Psi(t)}\bra{\Psi(t)}]$, where ${\rm Tr}_{\Gamma(\Xi)}$ indicates 
the partial trace over ${\cal H}_{\Gamma(\Xi)}$.

An observable $O_\Gamma$ of $\Gamma$ is most generally defined as - and 
identified with - a positive operator valued measure (POVM) on some 
measurable space that describes the possible measurement outcome of the 
observable itself. Multiplicative POVMs can be shown to coincide with
projection operator valued measures and, when acting on the real 
Borel space or a subset of it, define observables that will be hereafter 
dubbed {\it sharp} observables\cite{BuschLM96}.
Any such measure determines a unique Hermitian operator 
${\hat O}_\Gamma$ acting on ${\cal H}_\Gamma$, and viceversa. 
If $O_\Gamma$ is a {\it discrete} sharp observable, the spectral decomposition 
of the related operator reads
\begin{equation}
\hat{O}_\Gamma=\sum_{\gamma,i}\omega_\gamma\ket{\gamma i}\bra{\gamma 
i}~,
\label{e.OGamma}
\end{equation}
where $\{\omega_\gamma\}$ is the set of different $\hat O_\Gamma$-eigenvalues, 
with respective degeneracy ${\rm d}_\gamma$, the multiple index $\gamma 
i$ runs from $\gamma 1$ to $\gamma {\rm d}_\gamma$, and the $\hat 
O_\Gamma$-eigenvectors $\{\ket{\gamma i}\}$ form an orthonormal basis 
for ${\cal H}_\Gamma$. 

Further ingredients of a scheme designed for describing the measure of 
$O_\Gamma$ are {\it i)} a pointer observable $O_\Xi$ of $\Xi$, to be correlated 
with $O_\Gamma$, {\it ii)} a pointer function $f$ correlating the 
value sets of $O_\Xi$ and $O_\Gamma$, {\it iii)} a measurement coupling $V$
between $\Gamma$ and $\Xi$, ultimately responsible for the $\Psi$-state 
transformation $\rho_\Psi(0)\xrightarrow{V}\rho_\Psi(t)$
occurring during the preliminary stage of the process, 
i.e. before the actual production of a specific outcome is obtained.
In order to define a measurement scheme, a state transformation must 
feature some specific properties; most importantly it must 
guarantee that the probability reproducibility condition holds,
i.e. that 
\begin{eqnarray}
{\rm p}^{O_\Gamma}_{\rho_\Gamma(0)}(\omega_\gamma)&=&
{\rm p}^{O_\Xi}_{\rho_\Xi(t)}(f^{-1}(\omega_\gamma))\label{e.PRC}\\
\forall\omega_\gamma~{\rm in~the~set}~\{\omega_\gamma\}
&{\rm and}& {\rm for~all~possible~}\rho_\Gamma(0)~,
\nonumber
\end{eqnarray}
where p$^{O_\Gamma}_{\rho_{\Gamma}(0)}(\omega_\gamma)$ is the 
probability 
measure for the value $\omega_\gamma$ of $O_\Gamma$ when $\Gamma$ is in 
the state $\rho_\Gamma(0)$, and 
p$^{O_\Xi}_{\rho_\Xi(t)}(f^{-1}(\omega_\gamma))$
is that for the value
$f^{-1}(\omega_\gamma)$ of $O_\Xi$ when $\Xi$ is in the state 
$\rho_\Xi(t)={\rm Tr}_\Gamma[V[\rho_\Psi(0)]]$. 
Although discussing the meaning and relevance of condition (\ref{e.PRC}) 
goes beyond the purpose of this work, we notice the following:
The Shannon entropy associated to the distribution of probability 
measures p$^{O_\Gamma}_{\rho_\Gamma(0)}$ on the discrete set 
$\{\omega_\gamma\}$ 
is interpreted, from an information-theoretical viewpoint, as the 
average deficiency of information on the observable $O_\Gamma$ of 
$\Gamma$ before the measurement starts $(t=0)$. On the other hand, 
requiring that equality (\ref{e.PRC}) hold, ensures that the above 
deficiency equals the potential information gain upon measuring the 
observable $O_\Xi$ of $\Xi$, which corresponds to the naive 
notion that measuring implies information gain. In fact 
Eq.~(\ref{e.PRC}), with its left- and right-hand side referring to 
$\Gamma$ 
and $\Xi$, respectively, formally represents an information flow 
from the principal system to the apparatus: however, despite this 
information transfer occur whenever $V$ generates entanglement between 
$\Gamma$ and $\Xi$, in Sec.~\ref{s.ExtractInfo} we will show that decoherence 
has an essential role in guaranteeing that $t$ be such that the amount 
of information actually 
transferred, quantified by the above Shannon entropy, be different from 
zero. Notice that the distribution of probability 
measures p$^{O_\Xi}_{\rho_\Xi(t)}$ on the set $\{f^{-1}(\omega_\gamma)\}$ 
can be obtained by reconstructing the state $\rho_\Xi(t)$ by quantum 
tomography\cite{DarianoL-P01,DarianoMP00}, i.e. by determining the 
expectation values ${\rm Tr}[\hat Z^I_\Xi\rho_\Gamma(t)]$ of an 
appropriate set $\{Z^I_\Xi\}$ of observables on $\Xi$.
    
Getting back to the measurement scheme, it can be shown that a 
sufficient condition for a state transformation to qualify as a proper 
pre-measurement, by this meaning that it fulfils Eq.~(\ref{e.PRC}), is 
that $V$ be a trace-preserving linear mapping. When $V$ is further 
assumed to be unitary, the process coincides with the one first 
described by von Neumann\cite{Neumann1996}, later generalised by several 
authors\cite{LondonB39,Wigner52,ArakiY60,Yanase61,ShimonyS71,ShimonyS79,Busch87,BuschS89} and 
characterised in Ref.~\cite{Ozawa84} under the 
name of {\it conventional measuring 
process}.

The pre-measurement step of this type of process is defined by a unitary 
operator ${\hat U}$ on ${\cal H}_\Psi$ satisfying
\begin{equation}
\hat U(\ket{\gamma i}\otimes\ket{\Xi})=
\sum_j \ket{\gamma j}\otimes\ket{\Xi^\gamma}~,
\label{e.U}
\end{equation}
for any $\gamma,i$. Notice that, as we are not measuring the degeneracy 
parameter $i$, Eq.~(\ref{e.U}) allows the possibility that $\Gamma$ 
remain in any state of the $\hat O_\Gamma$-invariant subspace of 
${\cal H}_\Gamma$ corresponding to the eigenvalue $\omega_\gamma$.
If $O_\Xi$ is a sharp observable, with $\hat O_\Xi$ the corresponding 
hermitian operator, the propagator
\begin{equation}
\hat U_\tau\equiv e^{-i\frac{\tau}{\hbar}\hat{H}_\Psi}
\label{e.Ut}
\end{equation}
with 
\begin{equation}
\hat{H}_\Psi=g\hat{O}_\gamma\otimes\hat{O}_\Xi
+\hat{\mathbf 1}_{\Gamma}\otimes\hat{H}_\Xi~,
\label{e.H}
\end{equation}
where $\hat{H}_\Xi$ acts on 
${\cal H}_\Xi$ and $\hat {\mathbf 1}_\Gamma$ is the identity operator on 
${\cal H}_\Gamma$,
defines
a model, referred to as the {\it standard model} 
\cite{BuschLM96}, for properly
describing unitary pre-measurements as dynamical processes.

In what follows we will specifically study the standard model for the 
unitary pre-measurement of a discrete sharp observable and, for the sake 
of simplicity, we will further assume such observable to be 
non-degenerate. As for the parameter $\tau$ in Eq.(\ref{e.Ut}), we 
identify it with the time $t$.

Writing $\ket{\Gamma}$ in Eq.~(\ref{e.Psi0})
on the basis of the $\hat O_\Gamma$-eigenstates, from 
Eqs.~(\ref{e.Ut}-\ref{e.H}) it follows
\begin{equation}
\ket{\Psi(t)}=\sum_{\gamma}c_{\gamma}\ket{\gamma} 
\otimes\ket{\Xi^\gamma(t)}~
\label{e.Psit}
\end{equation}
at any time during the pre-measurement process, with
\begin{equation}
\ket{\Xi^\gamma(t)}\equiv
e^{-i\frac{t}{\hbar}\hat{H}^\gamma}\ket{\Xi}~,
\label{e.Xigammat}
\end{equation}
and $\hat H^\gamma\equiv g\omega_\gamma\hat O_\Xi+\hat H_\Xi$.
The density operator for $\Gamma$ consequently reads
\begin{equation}
\rho_\Gamma (t)=\sum_{\gamma}
|c_{\gamma}|^2\ket{\gamma}\bra{\gamma}+\sum_{\gamma\neq\gamma'}
c_{\gamma}c^*_{\gamma'}
\exval{\Xi^{\gamma '}(t)|\Xi^\gamma(t)}\ket{\gamma}\bra{\gamma '}~,
\label{e.rhoGammat}
\end{equation}
showing that, due to the structure of $\hat{H}_\Psi$, only the 
off-diagonal 
elements of the  above representation of $\rho_\Gamma$ evolve 
in time (which is why this type of evolution has been recently 
dubbed "off-diagonal dynamics"\cite{Calvani2013}).
It is of absolute relevance, as it will further result in
Sec.\ref{s.PRECS}, that the evolution of $\rho_\Gamma$ is exclusively
ruled by the time dependence of the overlaps
$\exval{\Xi^{\gamma'}(t)|\Xi^{\gamma}(t)}$.

\section{Off-diagonal dynamics by the PRECS}
\label{s.PRECS}
Our next step is that of obtaining an expression for $\rho_\Gamma(t)$ 
that allow us to go beyond the pre-measurement stage. To this aim we 
resort to the
parametric representation with environmental coherent states (PRECS):
the method has been recently introduced\cite{CalvaniEtal2013b}  as a 
tool for studying OQS 
with an environment that needs being considered quantum, but yet may 
have an extremely large Hilbert space.
It is based on the construction of generalised coherent 
states\cite{ZhangFG1990,Perelomov1972} for the 
environment, or environmental coherent states (ECS), relative 
to the group, usually referred to as "dynamical group", in terms of 
whose generators one can write the operators $\hat O_\Xi$ and 
$\hat{H}_\Xi$ in $\hat H_{\Psi}$.
Without entering into the details of their construction and 
properties\cite{Comberscue2012}, 
we recall that ECS,  hereafter indicated 
by $\ket{\Omega}$, form an overcomplete set 
on ${\cal H}_\Xi$ and are  in 
one-to-one correspondence with points $\Omega$ on a differentiable 
manifold ${\cal M}$. This correspondence is strictly local, but
coherent states are not orthogonal due to their overcompleteness, which 
suggests that they can be used also for studying pre-measurements where 
$O_\Xi$ is a POVM. On the other hand, and this is just one of the many 
ECS properties that make them the ideal tool for investigating the 
quantum to classical transition, their overlaps
exponentially vanish as dim${\cal H}_\Xi$ grows, and the manifold 
${\cal M}$ is demonstrated to be a proper phase-space in the classical 
limit\cite{Yaffe1982}.

The construction of ECS requires the (arbitrary) choice of a reference 
state $\ket{R}\in{\cal H}_\Xi$, whose representative point will define 
the origin of the reference frame on ${\cal M}$; the procedure entails 
the definition of an invariant (with respect to the dynamical group) 
measure $d\mu(\Omega)$ on ${\cal M}$, as well as a metric tensor ${\bf 
m}$. ECS provide an identity resolution on ${\cal H}_\Xi$ in the form
\begin{equation}
\hat{\mathbf 1}_{\mathcal{H}_\Xi}=\int_{\cal M}\, 
d\mu(\Omega)\ket{\Omega}\bra{\Omega}~.
\label{e.ECSidentityres}
\end{equation}

Due to their being constructed in relation to the dynamical group,
coherent states have peculiar dynamical properties, 
which are often summarised by the motto {\it "once 
a coherent state, always a coherent state"}\cite{ZhangFG1990}.
Referring to our specific setup, if the 
initial state of the apparatus is a coherent 
state, from 
\refx{e.Xigammat} it follows
\begin{equation}
\ket{\Xi^\gamma(t)}=e^{i\varphi^\gamma_t}\ket{\Omega^\gamma_t}~,
\label{e.Rgammat}
\end{equation}
with
\begin{itemize}
\item $\ket{\Omega^\gamma_t}$ the coherent state corresponding to the 
point $\Omega(t)$ on the trajectory on ${\cal 
M}$ defined by the solution of the classical-like equations of motion
\begin{equation}
i{\bf m}\frac{d\Omega}{dt}=\frac{\partial}{\partial\Omega^*}
H^\gamma(\Omega)~~~{\rm and~c.c.}~,
\label{e.eom}\
\end{equation}
with $H^\gamma(\Omega)=\exval{\Omega|\hat{H}^\gamma|\Omega}$,
\item
and 
\begin{equation}
\varphi^\gamma_t=
\int_0^t~
dy
\exval{\Omega^\gamma_y|\left(i\frac{\partial}{\partial y}-
\hat{H}^\gamma\right)|\Omega^\gamma_y}~.
\label{e.phi}
\end{equation}
\end{itemize}
Getting back to our composite system $\Psi$, once ECS are constructed, 
its subsystems $\Gamma$ and $\Xi$ can 
be formally split by inserting ${\mathbf 1}_{{\cal H}_\Xi}$
as from \refx{e.ECSidentityres} into any state $\ket{\Psi}$, including 
one written in the form (\ref{e.Psit}). In particular, 
choosing the initial state of the measuring apparatus as the reference 
state for the ECS construction, $\ket{R}=\ket{\Xi}$, and exploiting the fact that 
$d\mu(\Omega)$ is group-invariant, we can write
\begin{equation}
\ket{\Psi(t)}=\int_{\cal M}~d\mu(\Omega)
\chi_t(\Omega)\ket{\phi_t(\Omega)}\ket{\Omega}~,
\label{e.psipara}
\end{equation}
with 
\begin{eqnarray}
 |\phi_t(\Omega)\rangle&=&\frac{1}{\chi_t(\Omega)}
\sum_\gamma c_\gamma\exval{\Omega|R^\gamma_t}\ket{\gamma}~,
\label{e.phipara}\\
\chi_t(\Omega)&=&
\sqrt{\sum_\gamma|c_\gamma|^2h^\gamma_t(\Omega)}~,
\label{e.chi}\\
h^\gamma_t(\Omega)&=&|\exval{\Omega|R^\gamma_t}|^2~,
\label{e.hgammat}
\end{eqnarray}
where $\ket{R^\gamma_t}$ is the coherent state corresponding to the 
point $R^\gamma(t)$ on the trajectory defined by the solution of 
Eq.~(\ref{e.eom}) with initial condition $\Omega(0)=0$, 
and we have set
$\chi_t(\Omega)$ in ${R}^+$ by choosing its arbitrary phase 
equal to $0$. Due to $\exval{\Psi(t)|\Psi(t)}=1$ at any time, it is 
\begin{equation}
 \int_{\cal M} d\mu(\Omega)\chi^2_t(\Omega)=1~~~\forall t~.
\label{e.normchi}
\end{equation}
The above Eqs.(\ref{e.psipara}-\ref{e.hgammat}) define the parametric 
representation with environmental coherent states of $\ket{\Psi(t)}$.
It can be shown\cite{ZhangFG1990,CalvaniEtal2013b} that  corresponding 
form for 
$\rho_\Gamma(t)$ is 
\begin{equation}
\rho_\Gamma(t)=\int_{\cal M} d\mu(\Omega) \chi^2_t(\Omega)
\ket{\phi_t(\Omega)}\bra{\phi_t(\Omega)}~,
\label{e.rhopara}
\end{equation}
suggesting that $\chi^2_t(\Omega)$ can be interpreted, consistently with 
\refx{e.normchi}, as the density distribution of ECS on ${\cal M}$. To 
this respect it is worth noticing that 
$\chi^2_t(\Omega)=\exval{\Omega|\rho_\Xi(t)|\Omega}$.

\section{Extracting information from the apparatus: emergence of 
decoherence} 
\label{s.ExtractInfo} 
The description of the standard model for unitary pre-measurements of 
discrete, non degenerate, sharp observables by the PRECS essentially 
amounts to 
express $\rho_\Gamma(t)$, as from Eq.~(\ref{e.rhoGammat}), in the form 
(\ref{e.rhopara}), with
$\chi^2_t(\Omega)=\sum_\gamma|c_\gamma|^2h_t^\gamma(\Omega)$ and the 
initial state of the measuring apparatus chosen as reference state
for constructing ECS. In fact, by comparing 
Eqs.~(\ref{e.rhoGammat}) and (\ref{e.rhopara}), it might seem 
that we ended up with having overturned the dependencies with respect to the 
scheme (\ref{e.U}), as \refx{e.rhopara} 
shows that $|\phi_t(\Omega)\rangle$ depends on the 
environmental parameter $\Omega$, while the coherent state 
$\ket{\Omega}$ of the measuring apparatus is not
marked by the label ``$\gamma$''. This is because the signature of the 
interaction with $\Gamma$ is not in the ECS, 
that are defined independently of $O_\Gamma$, 
but rather in their density distribution $\chi_t^2(\Omega)$, which is 
where one should therefore look into, in order to extract information on 
$\Gamma$ via $\Xi$.

Let us now leave the 
pre-measurement process and consider the actual production of an 
outcome. Distinct coherent states, corresponding to distinct states 
of the measurement apparatus, will produce different outcomes, whose 
distribution will thus be associated with $\chi^2_t(\Omega)$. On the 
other hand, in order for this setup to produce an outcome with some 
informational content, it is necessary that the $\gamma-$components entering 
$\chi^2_t(\Omega)$, i.e. the terms $|c_\gamma|^2h_t^\gamma(\Omega)$, 
be sufficiently separated from each other to be distinguishable.
Aiming at formally expressing this condition, 
let us consider the functions $h^\gamma_t(\Omega)$ in \refx{e.hgammat}:
They are normalised distributions on ${\cal M}$ whose 
$\varepsilon$-support, defined 
as the region $S^\gamma_t\in {\cal M}$ such that  
$h^\gamma_t(\Omega)>\varepsilon~\forall \Omega\in S^\gamma_t$
(with $\varepsilon$ a reasonably small number in ${\mathbf R}^+$), moves on such 
manifold with time. 
If, after some time, we have 
\begin{equation}
S^\gamma_t\cap S^{\gamma '}_t=\emptyset~,
~\forall \gamma\neq\gamma'~,
\label{e.distinguishable}
\end{equation}
then each distribution $h^\gamma_t$ can be individually 
located, and a one-to-one correspondence between the label $\gamma$ and 
the region $S^\gamma_t$ on ${\cal M}$ is established. Notice that a 
weaker 
condition $(\cup_{i\in{\cal I}}
S^{\gamma_i}_t)\cap(\cup_{j\in{\cal J}}S^{\gamma'_j}_t)=\emptyset$ 
would also
establish a correspondence, although of a many-to-one type.
However, for the sake of clarity, in what follows we will concentrate 
upon the strong distinguishability condition (\ref{e.distinguishable}), 
identifying it with that
guaranteeing that an informative output can be extracted from the 
apparatus. This finally brings us to the question 
we aimed at answering:
how and why this condition, that somehow regards $\Xi$ only,  
is related with the occurrence of decoherence in the principal system 
$\Gamma$? In order to take this last step forward, consider 
\refx{e.rhoGammat}: If condition (\ref{e.distinguishable}) holds, 
we have
\begin{eqnarray}
\rho_\Gamma(t)&{\approx}&\sum_\gamma|c_\gamma|^2\int_{S^\gamma_t}~d\mu(\Omega)
h^\gamma_t(\Omega)\ket{\phi(\Omega,t)}\bra{\phi(\Omega,t)}\nonumber\\
&{=}&\sum_\gamma|c_\gamma|^2\ket{\gamma}\bra{\gamma}~,
\label{e.rhoGammaPRECSdeco}
\end{eqnarray}
that exactly expresses the vanishing of the off-diagonal elements of 
$\rho_\Gamma$ on the $\{\ket{\gamma}\}$ basis, i.e. the formal 
definition 
of decoherence for the principal system $\Gamma$.
This makes finally evident that decoherence is not one of the many 
byproducts of the measurement 
process, but rather a necessary condition for the configuration of the 
apparatus to embody some usable information on $\Gamma$. 

In order to better understand the construction that brought us to the 
above result, let us consider a simple example. Take $\Gamma$ as a 
quantum system with ${\rm dim}{\cal H}_\Gamma=2$ (usually referred to as 
qubit), and $\Xi$ as a single-mode bosonic field. Be the Hamiltonian
\begin{equation}
\hat{H}= 
g\sqrt{\hbar}\sigma^z(b+b^\dagger)+\nu b^\dagger b~,
\label{e.qubit-boson}
\end{equation}
with $[b,b^\dagger]=\hbar$,
$[\sigma^\alpha,\sigma^\beta] 
=i\varepsilon^{\alpha\beta\delta}{\sigma}^\delta$, 
$\alpha(\beta,\delta)=x,y,z$, and ${\Vec \sigma}$ the Pauli 
operator.
The above Hamiltonian is in the form~(\ref{e.H}) and the 
corresponding ECS, taking the reference state $\ket{R}$ such 
that $b\ket{R}=0$, are the usual field coherent states, with 
${\cal M}$ the complex plane, ${\rm d}\mu=d\Omega d\Omega^*/(\pi\hbar)$ 
the invariant measure, and ${\bf m}=\hbar^{-1}$ the (diagonal) metric tensor. 
Being ${\rm dim}{\cal{H}}_\Gamma=2$, the label $\gamma$ 
can only take two values, hereafter indicated by $\pm$, and the initial 
state $\Gamma$ can be written as $(c_+\ket{+}+c_-\ket{-})$, where 
$\ket{\pm}$ are the eigenstates of $\sigma^z$.
Solutions of \refx{e.eom} are two circles $R^\pm_t(\Omega)$ on the 
${\rm Re}(\Omega)-{\rm Im}(\Omega)$ plane,
passing through $(0,0)$, centred in $(\mp g/\nu,0)$, 
and gone through clockwise. 
\begin{figure}[t!]\centering
\includegraphics[scale=0.45]{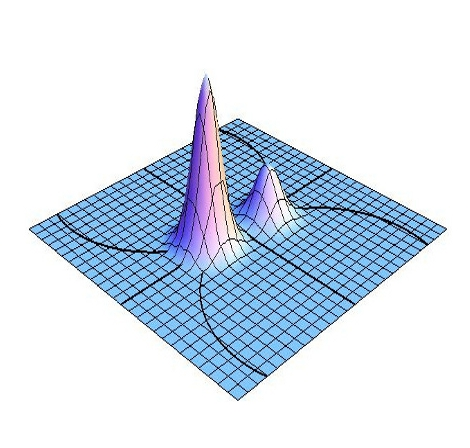}
\caption{\small 
Distribution $\chi^2_t(\Omega)$ at $t=\pi/30$ for the qubit-boson model, 
\refx{e.qubit-boson}, with $g\sqrt{\hbar}=2$ and $\nu=1$. In the 
initial state of the qubit it is $|c_+|^2=1/4$ and $|c_-|^2=3/4$}
\label{f.boson-snap}
\end{figure}
A snapshot of the ECS density distribution $\chi^2_t(\Omega)$ on the 
complex plane is shown in Fig.~\ref{f.boson-snap}, together with part 
of the orbits $R^\pm_t(\Omega)$: the two components $|c_\pm|^2 
h^\pm_t(\Omega)$ are already distinguishable, with the respective 
$\varepsilon$-supports $S^\pm_t$ quite well separated.
Notice that, despite this image portrays the environment
$\Xi$, it also mirrors the structure of the principal system state, 
$\rho_\Gamma$, due to the relation between condition 
(\ref{e.distinguishable}) and 
\refx{e.rhoGammaPRECSdeco}.
Although we have never mentioned it so far, it is worth noticing that 
the above relation between distinguishability of different $h^\gamma_t$ 
and diagonal form of $\rho_\Gamma(t)$ is established by the entanglement 
generation entailed by a non-trivial dynamical evolution of $\Psi$, 
such as that resulting from the interaction (\ref{e.H}). 

\begin{figure}[t!]\centering
\includegraphics[scale=0.45]{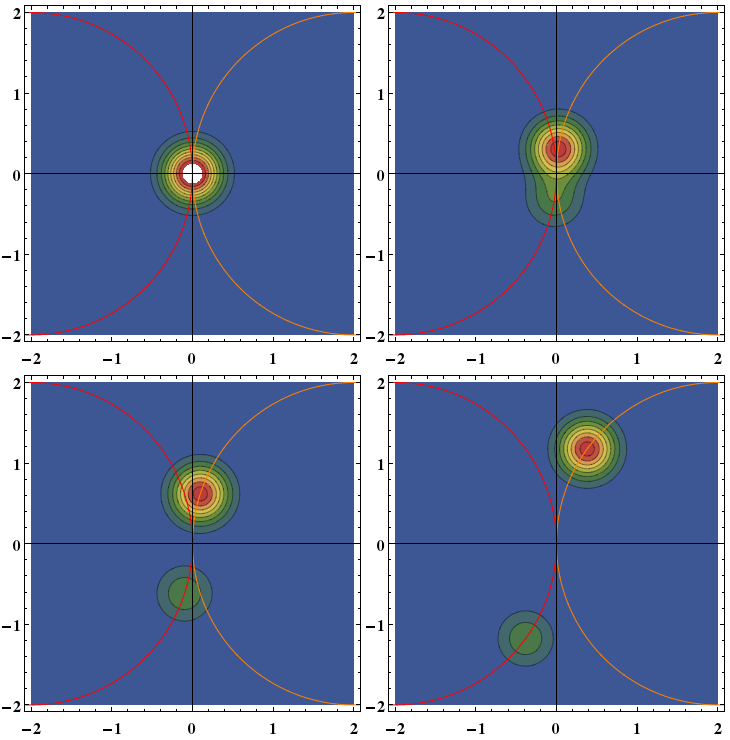}
\caption{\small 
Contour-plot of $\chi^2_t(\Omega)$ at
$t=0,~\pi/20,~\pi/10,~\pi/5$, (panels $A,~B,~C,~D$, resp.), with other 
parameters as in Fig.1~.
Different colours refer to different values of  $\chi^2_t(\Omega)$, 
increasing from blue ($\chi^2_t(\Omega)\sim0$) to white.}
\label{f.contour}
\end{figure}

The idea that the ECS distribution $\chi^2_t(\Omega)$ be the 
``image'' from which we can extract information on $\Gamma$ can be made more
precise by introducing the differential 
entropy\footnote{In information theory, it is the Shannon Entropy 
generalisation to continuous probability distributions.} for 
$\chi^2_t(\Omega)$
\begin{eqnarray}
 \mathcal{E}(t)&=&-\int_{\cal M} 
d\mu(\Omega)\chi^2_t(\Omega)\log\chi^2_t(\Omega)=\\
  &=&-\sum_\gamma\int_{\cal M} 
d\mu(\Omega)|c_\gamma|^2h_t^\gamma(\Omega)\log
\left(\sum_\gamma|c_\gamma|^2h_t^\gamma(\Omega)\right)~.
\end{eqnarray}
Referring to our example \refx{e.qubit-boson}, the explicit form of the 
distributions $h^\pm_t(\Omega)$ is 
\begin{equation}
h_t^\pm(\Omega)=\frac{1}{\pi\hbar}|\langle\Omega|R_t^\pm\rangle|^2=
  \frac{1}{\pi\hbar}e^{-\frac{1}{\hbar}|\Omega-R_t^\pm|^2}~,
\label{e.husimiboson}
\end{equation}
corresponding to gaussians centred in $R^\pm_t$ with constant variance 
$\hbar$. As the orbits $R^\pm_t$ initially coincide, it exists an early 
stage of the process, no matter the coefficients $c_\pm$,
during which $\chi^2_t(\Omega)$ keeps being an essentially
uni-modal distribution, centred in the origin of the complex plane, 
which implies $\mathcal{E}\sim \mbox{const}$, with no dependence on 
$c_\pm$, whatsoever.
On the other hand, the trajectories $R^\pm_t$ dynamically  
separate from each other and, after a certain time, condition 
(\ref{e.distinguishable}) starts holding, and the entropy
\begin{equation}
 \mathcal{E}\sim-\sum_{\gamma=\pm}|c_\gamma|^2\int_{S^\gamma_t}d\Omega 
d\Omega^* 
h_t^\gamma(\Omega)\log\left(|c_\gamma|^2h_t^\gamma(\Omega)\right)
 \label{e.entroqubitboson}
\end{equation}
is seen to depend on the coefficients $|c_\pm|^2$, and to quantify
the amount of information on the state $\Gamma$ that we can obtain 
adopting a measurement scheme based on the coupling
(\ref{e.qubit-boson}).
Fig.\ref{f.contour} offers a visual rendering of the above result
via the contour plot of $\chi^2_t(\Omega)$ on the complex plane  at 
different times: it is evident that the information content of the 
initial plot has nothing to do with the state of the principal system 
$\Gamma$, while 
the later emergence of two distinct fuzzy spots can be used to extract 
data on $c_+$ and $c_-$.

\section{Conclusions}
\label{s.Conclusions}

The analysis presented in this work
formally shows that the reason why decoherence of 
the principal system is a necessary ingredient of a significant 
measurement process is that the information content of the apparatus
would be otherwise null. To this respect it is important to recall that 
decoherence is defined as the dynamical process causing the vanishing of 
the off-diagonal elements of the system's density matrix, with respect 
to a precise basis on its Hilbert space. When considering the 
measurement process of a sharp observable, the relevant decoherence 
phenomenon is just that 
relative to the basis of eigenstates for the hermitian operator 
describing the observable itself. In fact, such basis 
explicitly comes into play when designing the interaction between 
principal system and apparatus, which is ultimately 
responsible for the information flow between the twos. Indeed, what does 
not depend on the specific observable to be measured, is the essential 
role of the dynamical entanglement generation, without which there would 
be no correlation between $\Gamma$ and $\Xi$ capable of leaving on the 
latter any trace of the quantum state of the former. This is an 
essential feature of quantum measurement, that accounts for the 
inability of approaches based on classical-like treatments of $\Xi$ to 
describe quantum measurements, as there cannot be entanglement between a 
quantum system to be observed and a classical apparatus that makes the 
measuring.

The necessary condition that both $\Gamma$ and $\Xi$ be quantum systems, 
on the other hand, raises another question worth being considered, 
namely whether one should expect coherence to be restored after a 
certain time or not. In fact, being $\Psi$ a quantum system, and its 
dynamics unitary, the evolution of both $\Gamma$ and $\Xi$ are 
superpositions of periodic motions, so that the time interval during 
which the apparatus is capable of conveying information is in 
principle finite. However, it can be 
shown\cite{Yaffe1982,Lieb1973} that when the 
dimension of the environmental Hilbert space grows, reflecting the fact 
that the apparatus is macroscopic, the environmental distributions 
$h^\gamma_t(\Omega)$ tend to Dirac $\delta$-functions and the recurrence 
time, i.e. the period of the unitary dynamics, diverges: as a 
consequence, 
decoherence occurs after an infinitesimally small time $\tau$, and 
coherence is never restored. To this respect, we underline that we have 
not considered the very last stage of the measurement process, where the 
Born's rule and the "wave-function collapse" come into play, and the 
unitarity of dynamics is lost. However, we believe the PRECS formalism, 
 can give 
original clues also with respect to these fundamental issues, but we 
postpone their possible analysis to future works.
Moreover, although we have restricted 
ourselves to a 
particular model that describes only a specific type of measurement, the 
present analysis might be useful also for treating more general
situations.

\section{Acknowledgements}
This work has been done in the framework of the {\it Convenzione 
operativa} between the Institute for Complex Systems of the italian 
National Research Council, and the 
Physics and Astronomy Department of the Univeristy of Florence.

\end{document}